\documentclass{article}

\usepackage{arxiv}
\usepackage[normalem]{ulem}
\usepackage[utf8]{inputenc} 
\usepackage[T1]{fontenc}    
\usepackage{hyperref}       
\usepackage{url}            
\usepackage{booktabs}       
\usepackage{amsfonts}       
\usepackage{nicefrac}       
\usepackage{microtype}      
\usepackage{lipsum}
\usepackage{multicol}
\usepackage[backend=biber]{biblatex}
\usepackage{graphicx}
\usepackage{amsmath}
\usepackage{esdiff}
\usepackage{xcolor}
\usepackage{siunitx}
\graphicspath{ {./images/} }

\newcommand{\Lagr}{\mathcal{L}}

\newcommand{\nunit}{n_\text{subunit}}
\newcommand{\nstripe}{n_\text{stripe}}

\title{Folding-Driven Auxetic Weft Knit Textiles with Integrated Capacitive Sensing}

\author{
 Kausalya Mahadevan \\
  John A. Paulson School of Engineering and Applied Sciences\\
  Harvard University\\
  Cambridge, MA 02138 \\
\And
 Helen E. Read \\
  John A. Paulson School of Engineering and Applied Sciences\\
  Harvard University\\
  Cambridge, MA 02138 \\
\And
 Anya X. Zhang \\
  John A. Paulson School of Engineering and Applied Sciences\\
  Harvard University\\
  Cambridge, MA 02138 \\
  \AND
   Louis-Justin Tallot \\
  John A. Paulson School of Engineering and Applied Sciences\\
  Harvard University\\
  Cambridge, MA 02138 \\
  \And
   Michelle C. Yuen \\
  Department of Mechanical \& Industrial Engineering\\
  Montana State University\\
  Bozeman, MT 59717\\
  \And
   Katia Bertoldi \\
  John A. Paulson School of Engineering and Applied Sciences\\
  Harvard University\\
  Cambridge, MA 02138 \\ \\
  \texttt{bertoldi@seas.harvard.edu} \\
}

\begin{document}
\maketitle
\begin{abstract}
Machine knitting provides a scalable platform for manufacturing multifunctional textiles in which geometry, mechanics, and embedded functionality can be programmed at the stitch level. However, predictive design tools capable of linking knit architecture to large-deformation mechanical response remain limited. Here, we develop a reduced-order spring-network model that captures the relaxation, unfolding, and deformation of knitted fabrics composed of checkerboard arrangements of rib and garter patches. The model accurately predicts the corrugated relaxed configuration of the knits and the evolution of local deformations under tensile loading using only linear extensional and torsional springs. Combining simulations with experiments, we show that the programmed unfolding of the corrugations generates tunable auxetic behavior, with both the magnitude of the negative Poisson's ratio and the strain at which it occurs governed by the unit-cell geometry. We further   integrate capacitive strain sensing directly during fabrication through partial plating of conductive yarns, eliminating post-processing. The resulting knitted capacitors exhibit programmable tradeoffs between strain sensitivity and sensing range, enabling either highly sensitive sensors over narrow deformation windows or lower-sensitivity sensors capable of measuring larger strains. Together, our modeling framework and fabrication strategy provide a route toward the rational design of mechanically programmable, sensorized knits with tailored shape-morphing and sensing functionalities.
\end{abstract}
\twocolumn

\section{Introduction}

Weft-knit textiles, composed of a grid of interlocking slip knots stabilized by neighboring loops, represent a long-established technology that has served as a primary interface between the human body and its environment. Owing to their lightweight and compliant nature, there has been growing recent interest in “smart” textiles, in which robotic components such as sensors, actuators, and circuitry are integrated directly into the fabric through embedded bladders and conductive elements for wearable and robotic applications \cite{sanchez2021textile, elmoughni2021machine,du2024haptiknit,Atalay2018ALayer,abbas_Tuning_2026}.

Machine knitting is an attractive fabrication approach due to its ability to integrate multiple materials (including active components) at the yarn level \cite{kilic2021omnifiber}, as well as to locally program stich patterns to produce complex, non-homogeneous 3D shell geometries such as branched tubular structures \cite{Narayanan2018AutomaticMeshes, Underwood2009Fix}.   As a result, knitting has enabled the development of both inflatable \cite{sanchez2023knitting,elmoughni2021machine,luo2022digital} and shape-memory actuators \cite{eschen2019performance,Eschen2020materialia,Granberry2017activeorthostaticfix,granberry2019functionally,granberry2021kinetically}.
By carefully arranging and orienting stitches,  tunable shape-changing responses and even  sensing capabilities have been demonstrated into inflatable actuators \cite{luo2022digital,sanchez2023knitting}. In addition, complex corrugated architectures can be realized by knitting sections on the front and back beds of the machine \cite{sanchez2023knitting, mahadevan2024knitting,niu2025geometric}. These corrugations lead to unique mechanical responses and shape transformations, enabling functionalities such as multistability and programmable shape changes \cite{sanchez2023knitting, mahadevan2024knitting}.

To advance knitting design and fully explore its capabilities, there is a need for models that can  reproduce the mechanical response of knitted structures and, in turn, enable predictive and optimized design. Existing modeling approaches for textiles span multiple scales, from simulating their behavior at the yarn level including interactions  of individual stitches \cite{ding2024unravelling,tajiri2025curling,Kaldor2008SimulatingLevel}, to homogenized representations of stitch-level mechanics \cite{du2024haptiknit,singal2024programming}, anisotropic sheet models \cite{nguyen2020design,mahadevan2024knitting}, empirical analytical expression to capture behaviors at the device scale \cite{o2022unfolding}.  Each of these methods has their own drawbacks, and can require the fitting or measuring of many parameters, including friction, bending stiffness, and anisotropic extensional stiffnesses. 

In this work, we focus on textiles composed of a checkerboard pattern of rib and garter patches, which gives rise to complex 3D corrugated geometries \cite{salmon2025structural,ma2017review}. We develop a modeling framework based on a network of linear torsional and extensional springs to approximate their nonlinear behavior. We show that this model captures both the geometry and kinematic response of the considered textiles. 
Combining simulations with experiments, we characterize their kinematic response as the geometric parameters of the checkerboard pattern are varied. 
We find that stretching induces pronounced internal rotations, giving rise to auxetic behavior. 
Finally, we integrate capacitive sensing capabilities into these fabrics and investigate how the initial corrugated configuration and resulting kinematics affect their sensing performance.

\section{Response of rib and garter fabrics}
\begin{figure}
  \includegraphics{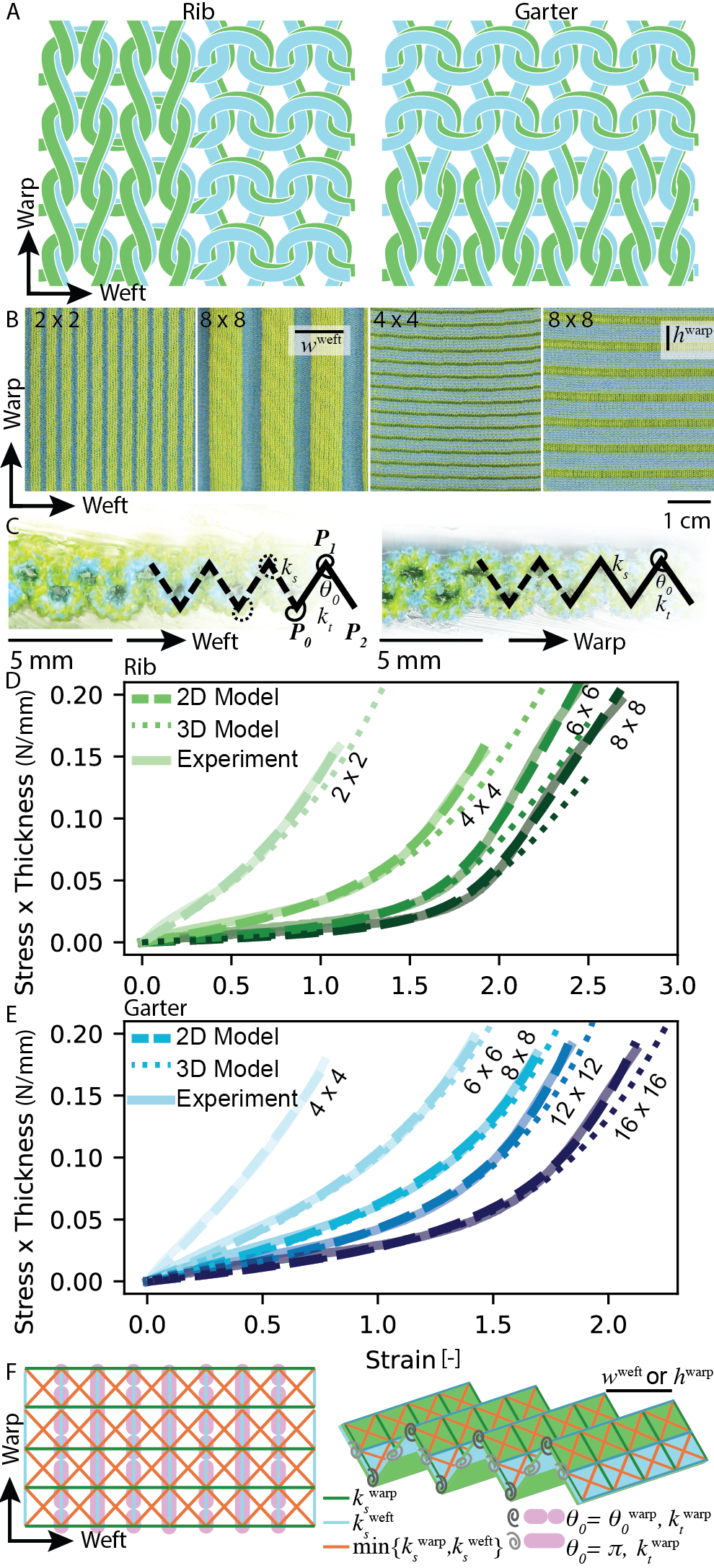}
  \caption{ A) Knitting diagrams of rib and garter stitch. 
  B) Top view of 2$\times$2 and 8$\times$8 rib fabrics and 4$\times$4 and 8$\times$8 garter fabrics under finite strain.  C) Cross-section photos of 4 $\times$ 4 rib and 8 $\times$ 8 garter fabrics. The   corrugated patterns  can be described with a network of springs. D-E) Experimentally measured and numerically predicted stress-strain curves for D)  $n \times n$ rib and E)  $n \times n$ garter fabrics. The results of both the  2D and 3D model are shown.    F) We can expand the  spring model to 3D to  capture the response of fabrics comprising patterns of rib and garter in
2D.}
  \label{fig: rib x garter}
\end{figure}
Rib $n \times n$ and garter $n \times n$ fabrics consist of vertical and horizontal stripes of front and back bed knits $n$ stitches wide \cite{sanchez2023knitting, mahadevan2024knitting} (Fig. \ref{fig: rib x garter}A). 
Plating two colors of yarn allows us to to visualize the technical front and reverse of the fabric as two different colors \cite{mahadevan2024knitting} (Fig. \ref{fig: rib x garter}B). 
As discussed by several studies \cite{niu2025geometric,singal2024programming,ding2024unravelling,sanchez2023knitting,mahadevan2024knitting}, the internal structure of stitches in rib and garter knits induces curvature, which gives rise to characteristic parallel ridges. We observe these ridges in the cross-section of these textiles in Fig.~\ref{fig: rib x garter}C. The relaxed curved shape of these knit patterns depends on many factors, including, but not limited to, the yarn choice, tension during knitting, and contact between the ridges. We measure the mechanical response to uniaxial load in an Instron tensile testing machine, and always test perpendicular to the rib or garter parallel ridges. Under uniaxial tension these corrugations unfold. Once the textile is flat,  the stitches and yarns themselves begin to stretch, leading to the typical J-shaped mechanical response \cite{sanchez2023knitting}. 
When we keep all other factors the same, we observe the mechanical response depends on $n$, the stripe width (Figs.~\ref{fig: rib x garter}D-E). In particular, we observe that  lower $n$ leads to  more tightly folded ridges and, consequently,    initial higher stiffness and earlier stiffening. 
For $2\times2$ rib and $4\times4$ garter the J-shaped mechanical response is not yet visible. 
On this scale, the stripe width and stitch width are too similar, and the mechanics of the individual stitch and the unfolding of the textile cannot be separated. 

\section{Modeling knit folds as a network of linear springs}
To capture the response of the knits as a function of $n$, we develop a simplified model in which the corrugations are represented as a zigzag arrangement of linear extensional springs connected by linear torsional springs. 
We begin with pure rib or garter knits, where the textile is corrugated in a single direction. In this case, we model the system as a one-dimensional array of springs (Section \ref{2Dmodel}). We then extend the approach to non-periodic rib and garter patterns in two dimensions by generalizing the model to a two-dimensional network of springs (Section \ref{3Dmodel}).

\subsection{2D model of rib or garter}\label{2Dmodel}
For a textile corrugated in a single direction (rib or garter), we adopt a plane strain assumption with zero strain along the corrugations.
The corrugations are modeled as a zigzag arrangement of linear extensional springs, with a peak-to-peak distance $w^\text{weft}$ for the rib and $h^\text{warp}$ for the garter, forming an angle $\theta_0$ between adjacent segments, and connected by linear torsional springs at the joints (Fig.~\ref{fig: rib x garter}C). The mechanical response of the system is governed by four parameters: the extensional spring stiffness $k_s$, the torsional spring stiffness $k_t$, the peak-to-peak distance $w^\text{weft}$ for the rib and $h^\text{warp}$ and the rest angle $\theta_0$ of the torsional spring connecting adjacent extensional elements.

The total energy of such spring network is given by
\begin{equation}\label{eq_en_springs}
E_{\text{tot}} =
\frac{1}{2} k_s \sum_i \Delta \ell_{0i}^2 
+
\frac{1}{2} k_t \sum_i \Delta \theta_{0i}^2,
\end{equation}
where $\Delta \ell_{0i}$  denotes the change in  length of the $i$-th extensional spring and $\Delta \theta_{0i}$  represents the change in  angle of the $i$-th rotational spring. 

To simulate stretching, we fix the leftmost node of the spring network and prescribe a displacement in the $x$-direction at the rightmost node. Under these boundary conditions, the equilibrium positions of all intermediate nodes, and thus the configuration of the entire structure, are obtained by minimizing the total elastic energy using \texttt{scipy.minimize}. Once the energy-minimizing configuration is found for a given imposed displacement, the reaction force at the rightmost node is computed. Repeating this procedure for different levels of applied displacement yields the stress--strain response of the spring system for given values of $k_t$, $k_s$, $w^\text{weft}$ or $h^\text{warp}$ and $\theta_0$.
For each value of $n$, the stiffnesses $k_s$ and $k_t$ and rest angle $\theta_0$ are determined by fitting to the corresponding experimental stress--strain curves of rib and garter $n \times n$ samples (Figs.~\ref{fig: rib x garter}D and E), while   $w^\text{weft}$ and $h^\text{warp}$ are obtained from measurements of the corrugation geometry in the experimental samples (Fig.~\ref{fig: rib x garter}B).  The resulting fitted parameters are reported in Tables~\ref{tab:garter vals} and~\ref{tab:rib vals}.
\begin{table}[]
\centering
\caption{Fitted parameters for the garter stitch model as a function of system size $n \times n$. The parameters $k_s^{\text{warp}}$ and $k_t^{\text{warp}}$ and $\theta_0^{\text{warp}}$ are obtained by fitting to experimental stress--strain data, while $h^{\text{warp}}$ and $w^{\text{weft}}$ are the measured periodicity of the stripe.}
\label{tab:garter vals}
\begin{tabular}{l|llll}
$n \times n$ & $h^{\text{warp}} [\si{mm}]$ & $k_s^{\text{warp}}[\si{N/mm}]$ & $\theta_0^{\text{warp}}[\si{radians}]$ & $k_t^{\text{warp}}[\si{N}]$ \\ \hline
4            & 1.03  & 4.24  & 0.76       & 0.144 \\
6            & 1.21  & 4.8   & 0.7        & 0.071 \\
8            & 1.44  & 4.78  & 0.68       & 0.045 \\
12           & 2.03  & 1.78  & 0.74       & 0.025 \\
16           & 2.6   & 1.16  & 0.71       & 0.018
\end{tabular}
\end{table}

\begin{table}[]
\centering
\caption{Fitted parameters for the rib stitch model as a function of system size $n \times n$. The parameters $k_s^{\text{weft}}$ and $k_t^{\text{weft}}$ and $\theta_0^{\text{weft}}$ are obtained by fitting to experimental stress--strain data, while $h^{\text{warp}}$ and $w^{\text{weft}}$ are the measured periodicity of the stripe.}
\label{tab:rib vals}
\begin{tabular}{l|llll}
$n \times n$ & $w^{\text{weft}}[\si{mm}]$ & $k_s^{\text{weft}}[\si{N/mm}]$ & $\theta_0^{\text{weft}}[\si{radians}]$ & $k_t^{\text{weft}}[\si{N}]$ \\ \hline
2            & 1.09  & 1.825  & 0.874       & 0.0498 \\
4            & 1.41  & 0.817   & 0.772        & 0.0178 \\
6            & 1.93  & 0.562  & 0.754       & 0.0072 \\
8           & 2.44  & 0.455  & 0.732       & 0.0048 
\end{tabular}
\end{table}
As shown in Fig.~\ref{fig: rib x garter}D-E this simplified model is able to capture the stress-strain response of rib and garter stitches for all considered values of $n$, allowing us to describe the mechanical response of these complex curved knits under extension. 
\subsection{3D model}\label{3Dmodel}
To extend the model to non-periodic patterns of rib and garter in 2D, we  generalize to a two-dimensional network of springs.  This network consists of a rectangular array of springs with diagonal cross-braces (Fig.~\ref{fig: rib x garter}F). This can be thought of as an extrusion of the 2D model in Fig.~\ref{fig: rib x garter}C. The rotational springs now act over two faces that share a single edge (Fig. \ref{fig: rib x garter}F).
All springs parallel to the weft direction have a stiffness $k_s^{\text{weft}}$, determined by the fitting in Fig. \ref{fig: rib x garter}D-E and reported in Tables \ref{tab:garter vals} and \ref{tab:rib vals}. 
The springs parallel to the warp direction have stiffness $k_s^{\text{warp}}$.

Each rectangle of springs is cross-braced by two diagonal springs.
To reduce the total amount of extensional stiffness, the diagonal spring stiffness is the smaller of the rib and garter stiffness, $k_s^\text{diag} = \text{min} \{k_s^\text{warp}, k_s^\text{weft}\}$.
This extension to three dimensions is illustrated in Fig.~\ref{fig: rib x garter}F.

The total elastic energy of the system is given by
\begin{equation}\label{eq22}
V = E_{\text{tot}} + V_{\text{wall}},
\end{equation}
where $E_{\text{tot}}$ is defined in Eq.~(\ref{eq_en_springs}) and $V_{\text{wall}}$ is a quadratic penalty term that prevents large-scale out-of-plane buckling by bounding the out-of-plane displacement of each node:
\begin{equation}
- z_{\max} \le z_i \le z_{\max},
\qquad \text{with } z_{\max} = 6.0~ \si{mm}.
\end{equation}
Here, $z_i$ denotes the current $z$-coordinate of node $i$. We define $V_{\text{wall}}$ as
\begin{equation}
V_{\mathrm{wall}}
=
\frac{1}{2}\,k_{\mathrm{wall}}
\sum_{i=1}^{N}
\left[
(\delta_i^{+})^2 + (\delta_i^{-})^2
\right],
\end{equation}
where $k_{\mathrm{wall}} = 10^2$ \si{N/mm} and
\begin{equation}
\delta_i^{+} = \max(z_i - z_{\max}, 0), 
\qquad
\delta_i^{-} = \max(-z_{\max} - z_i, 0).
\end{equation}
This quadratic penalty term serves to maintain approximately planar deformations while avoiding convergence and computational issues.

The Lagrangian of the model is then  written as
\begin{equation}
\Lagr = T - V,
\label{eq:lagrangian}
\end{equation}
where $V$ is given by Eq. (\ref{eq22}) and $T$ denotes the kinetic energy of the system
\begin{equation}
T(\dot{\mathbf{u}})
=
\frac{1}{2}
\sum_{i=1}^{N}
m_i
\left\|
\dot{\mathbf{u}}_i
\right\|^2,
\label{eq:kinetic_energy_nodal}
\end{equation}
where $\dot{\mathbf{u}}_i$ and $m_i$ are the velocity ad concentrated mass of node $i$ and $N$ is the total number of nodes. 
The lumped mass $m_i$ is defined as
\begin{equation}
m_i = \rho_s \, A_i
\quad \text{with} \quad
A_i \approx \frac{w^{\text{weft}}}{4} \, \frac{h^{\text{warp}}}{4},
\end{equation}
where $\rho_s = 0.11 \si{ kg/m^2}$ is a constant surface density across our textile and $A_i$ is the effective nodal area.

Finally, we take advantage of automatic differentiation (using \texttt{Jax.grad})  to take the partial derivatives of $\Lagr$ with respect
to all degrees of freedom of the system and obtain the
equations of motion as

\begin{equation}
    \frac{\mathrm{d}}{\mathrm{d}t} \left(\frac{\partial \Lagr}{\partial \mathbf{\dot{u}}}\right) - \frac{\partial \Lagr}{\partial\mathbf{u}} + c \mathbf{\dot{u}} = 0,
    \label{eq:lagr_eq_of_motion}
\end{equation}
where $c$ is an effective viscous damping parameter that we use to account for for viscous and frictional dissipation in the fabrics.
Equation (\ref{eq:lagr_eq_of_motion}) is a highly nonlinear system of ordinary differential equations that we numerically solve for the nodal positions and velocities using \texttt{diffrax.Tsit5}, a 5th order explicit Runge--Kutta method.

We verify this model by investigating a textile sample of purely rib or garter corrugations. 
For a rib textile sample, we set up a grid of size $w^\text{weft}/4$. In accordance with the corrugations of the rib sample, the edges are populated with rotational springs with a stiffness $k_t^\text{weft}$ and alternating rest angles of $\pi$ and $\theta_0^\text{weft}$. This spring system is shown in 2D and 3D in Fig. \ref{fig: rib x garter}F and values from Table  \ref{tab:rib vals}) are assigned to each spring. For a garter textile, the spring system is set up the same way, but using values from Table \ref{tab:garter vals}. 

We then apply a fixed displacement to the edge of our 3D rib or garter model and calculate the reaction forces. The normalized stress-strain response for the whole family of rib and garter corrugations is shown in Figs. \ref{fig: rib x garter}D-E. When applied to pure rib and garter fabrics, the model reproduces the same results as the two-dimensional formulation (Figs.~\ref{fig: rib x garter}D–E).

\section{Relaxation of a pattern of rib and garter patches}

\begin{figure}[h]
  \centering
  \includegraphics{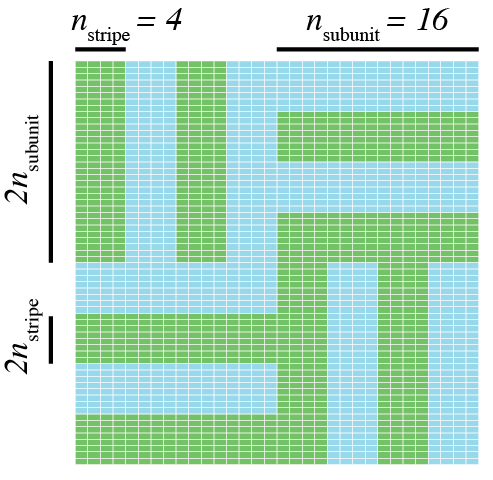}
  \caption{An example schematic of the unit cell of the fabric considered in this study, with $(n_{\text{stripe}}, n_{\text{subunit}})=(4,16)$. The white-outlined rectangles represent the bounding boxes of individual stitches and have an aspect ratio of 1:2. }
  \label{fig: knit stitches}
\end{figure}

We can use this framework to pattern arbitrary combinations of rib and garter patches. In this work, we focus on a checkerboard arrangement specified by two parameters: $n_{\text{stripe}}$, the number of stitches across the width of a rib stripe and $n_{\text{subunit}}$, the number of stitches along the length of a garter stripe.
Because the aspect ratio of individual knit stitches is approximately $1:2$ (Fig. \ref{fig: knit stitches}), the number of stitches across the width of a garter stripe is $2n_{\text{stripe}}$, and the number of stitches along the length of a rib stripe is $2n_{\text{subunit}}$ (Fig.~\ref{fig: knit stitches}).
Accordingly, in the model, the spring stiffness associated with an $n \times n$ rib fabric is paired with the spring stiffness associated with a $2n \times 2n$ garter fabric. 

\begin{figure}[h]
  \includegraphics{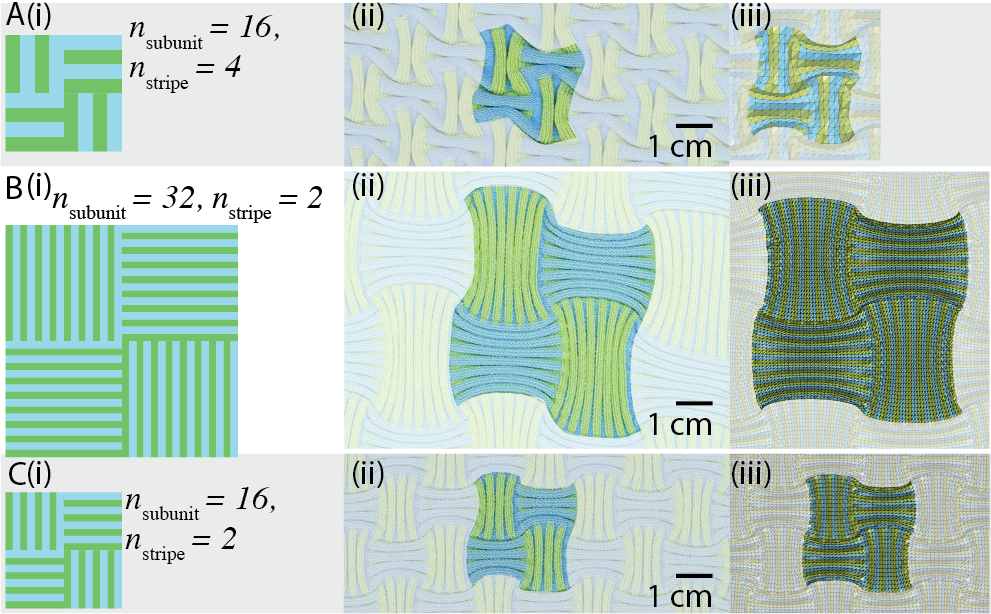}
  \caption{A--C) Relaxed configurations of fabrics composed of alternating rib and garter patches with $(n_{\text{stripe}}, n_{\text{subunit}})=$ A) $(4,16)$, B) $(2,32)$, and C) $(2,16)$. For each fabric, we show (i) a schematic of the unit cell, (ii) an experimental image of the relaxed configuration, and (iii) the corresponding numerically predicted relaxed configuration.}
  \label{fig: unit cells}
\end{figure}

On the knitting machine, stitches are formed on either the front or back bed and are transferred between the two when transitioning between patches. 
The yarn is held under tension as it is fed into the machine, and each newly formed loop is tensioned by a combination of needles, pressers, and rollers. 
Once the textile is removed from the machine, it relaxes into a corrugated shape, which we further enhance through the application of steam. In Fig.~\ref{fig: unit cells}A(ii), we show the relaxed and corrugated configuration for a  fabric with $n_{\text{stripe}} = 4$ and $n_{\text{subunit}} = 16$. 
 We observe that the center of each rib or garter patch contracts more than the edges, producing a sinusoidal boundary between adjacent patches and inducing rotation both at the center and along the boundaries of each unit cell. This contraction is present across all samples; however, the amplitude of the sinusoidal boundary between rib and garter patches depends on $n_{\text{stripe}}$ and $n_{\text{subunit}}$ (Figs.~\ref{fig: unit cells}B-C (ii)).  

\begin{figure}[h]
  \includegraphics{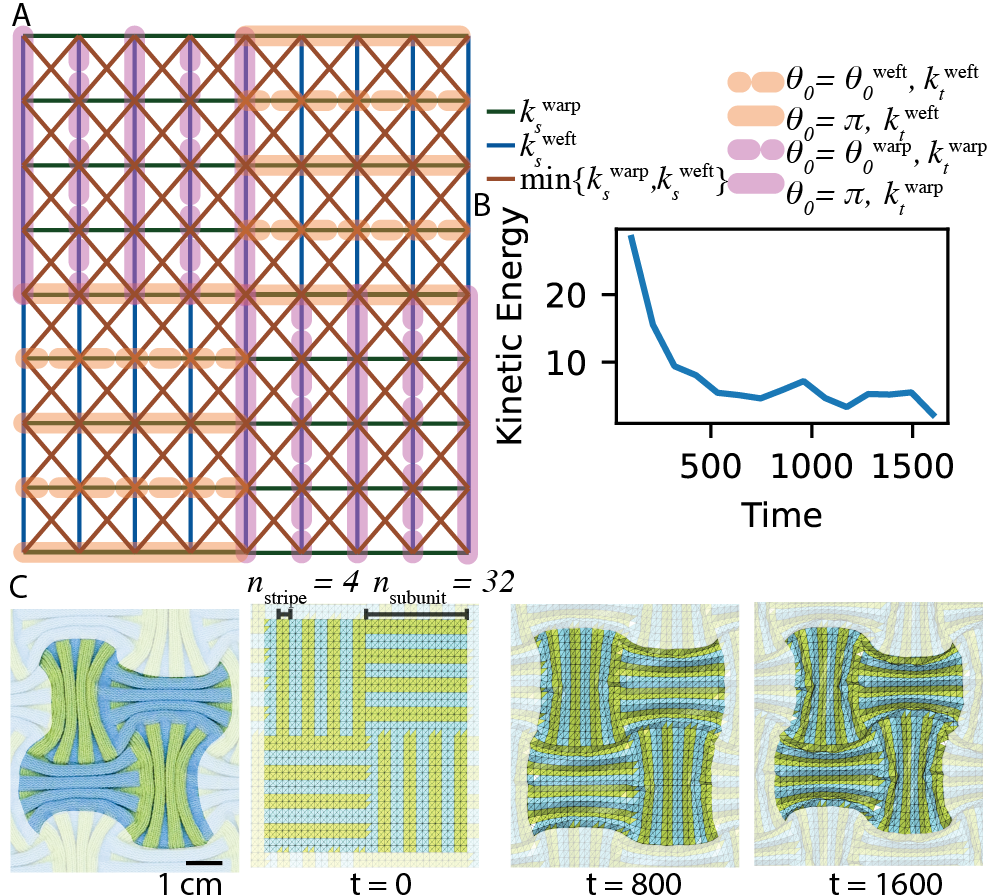}
  \caption{A) Spring network used to model a fabric composed of alternating rib and garter patches. B) Evolution of the kinetic energy during relaxation of the spring network from an initially flat configuration to its equilibrium corrugated state for a fabric with $(n_{\text{stripe}}, n_{\text{subunit}})=(4,32)$. C) Numerical snapshots of the fabric during the relaxation process.}
  \label{fig: unit simulation}
\end{figure}

The relaxed 3D shape of these fabrics can also be predicted using the model.  To create these geometries using our spring model, we again create a planar grid of springs (Fig.~\ref{fig: unit simulation}A), where the size of each rectangle is $w^\text{weft}/4$ and $h^\text{warp}/4$ and assign stiffness to each spring according to the values reported in Tables \ref{tab:garter vals} and \ref{tab:rib vals}.
Analogous to the individual stitches being released from the needle hooks and relaxing into a final textile configuration, we allow our spring model to relax from its flat configuration into its corrugated state. As the system approaches its equilibrium state, the kinetic energy tends to zero (Fig. \ref{fig: unit simulation}B). Each simulation is run for a fixed time of 1600 steps. 
The numerically predicted relaxed configuration exhibits frustration at the interfaces between rib and garter patches, inducing curvature and giving rise to sinusoidal boundaries between neighboring subunits (Fig.~\ref{fig: unit simulation}C). The sinusoidal boundaries predicted by the simulations compare well with those observed experimentally (Figs.~\ref{fig: unit cells}A--C(iii)) and exhibit amplitudes that depend on both $n_{\text{stripe}}$ and $n_{\text{subunit}}$.

\begin{figure}[t]
    \centering
    \includegraphics{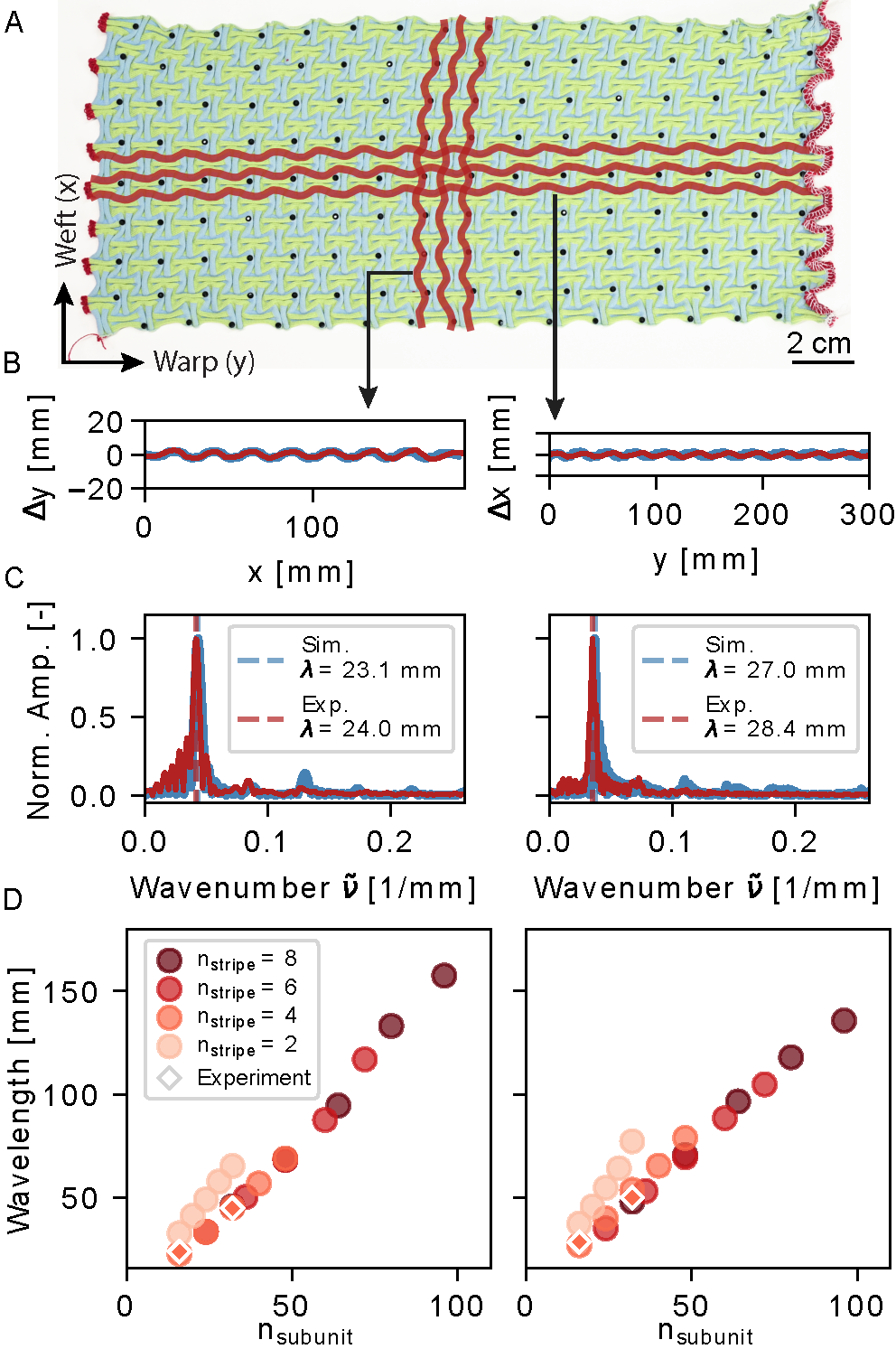}
    \caption{A) Top-view image of the relaxed configuration of a fabric with $(n_{\text{stripe}}, n_{\text{subunit}})=(4,16)$. The boundaries of the centrally located unit cells along the $x$- and $y$-directions are highlighted in red. B) Comparison of the unit-cell boundaries along the $x$- and $y$-directions as measured experimentally and predicted by the model. Boundaries along the $x$-direction are shown on the left, while those along the $y$-direction are shown on the right. C) Discrete Fourier transform (DFT) of the unit-cell boundaries. The characteristic wavelength is extracted from the location of the dominant spectral peak. D) Characteristic wavelengths extracted from the DFT analysis for all simulated fabrics; experimental measurements are shown as diamonds.}
    \label{fig:wavelength_analysis}
\end{figure}

Next, we quantitatively compare the relaxed corrugated configurations observed in our samples with those predicted by the simulations. To this end, we trace the unit cell boundaries along the samples in both the warp ($y$) and weft ($x$) directions (Fig.~\ref{fig:wavelength_analysis}A) and extract the relaxed nodal coordinates from simulations of the same geometry. Sample-scale deformation is removed with a Savgol filter, and we compare the filtered unit cell boundaries in Fig.~\ref{fig:wavelength_analysis}B.
To compare the extracted curves, we perform discrete Fourier transforms (DFTs) and calculate the wavenumber $\tilde{\nu}_{\text{peak}}$ corresponding to the maximum peak. 

The wavelength of the unit cell boundaries is then given by $\lambda = 1/\tilde{\nu}_{\text{peak}}$. For the geometry with $\nstripe = 4$ and $\nunit = 16$ shown in Fig.~\ref{fig:wavelength_analysis}A, the simulations and experiments exhibit the same wavelength along both directions, with agreement within approximately 1 mm (Fig.~\ref{fig:wavelength_analysis}C).
We repeat this analysis for $\nstripe \in {2,4,6,8}$ and multiple values of $\nunit$. Specifically, for $\nstripe = 2$, we consider $\nunit \in [4,8]\times 2\nstripe$, whereas for $\nstripe > 2$, we consider $\nunit \in [2,6]\times 2\nstripe$. The results, reported in Fig.~\ref{fig:wavelength_analysis}D, reveal an approximately linear relationship between $\nunit$ and the calculated unit-cell wavelength for all values of $\nstripe$. Furthermore, for $\nstripe > 2$, the data collapse onto a common slope, indicating a universal scaling relationship. These observations support the intuitive notion that the number of stitches per unit cell determines the wavelength, while the proportionality constant is set by the stripe width $\nstripe$.
The fabrics with  $\nstripe = 2$ exhibits the largest deviation from this common trend. We attribute this behavior to the absence of a distinct soft region in the $2\times2$ rib and $4\times4$ garter structures shown in Fig.~\ref{fig: rib x garter}D,E, which alters the effective mechanics governing unit-cell formation.

\section{Kinematics under tension of the checkerboard pattern}

We expect the frustration, curvature, and rotation observed in our relaxed samples to lead to auxetic behavior under tensile loading \cite{Hu2011DevelopmentTechnology,Hu2019AuxeticTextiles, Alderson2012AuxeticStructures}.
To characterize the tensile response of these knit structures, we fabricate samples approximately 200 mm wide and 400 mm long. The exact width is set to be as large as possible on the 360 needles while preserving an integer number of unit cells. These samples are tested in an Instron uniaxial tensile testing machine; to reduce out-of-plane deformation, the samples are sandwiched between two clear acrylic plates. 
To track deformation and set the samples away from the plates, plastic hemispheres are glued to the corners of each unit cell. 
As we stretch the sample, we track the locations of these markers \cite{bordiga2023humble} and use their positions to calculate the local deformation gradient $\mathbf{F}^i$ of the $i$th unit cell \cite{shan2015design}.
This allows for the calculation of local strains and Poisson's ratio. 

\begin{figure}
    \centering
    \includegraphics{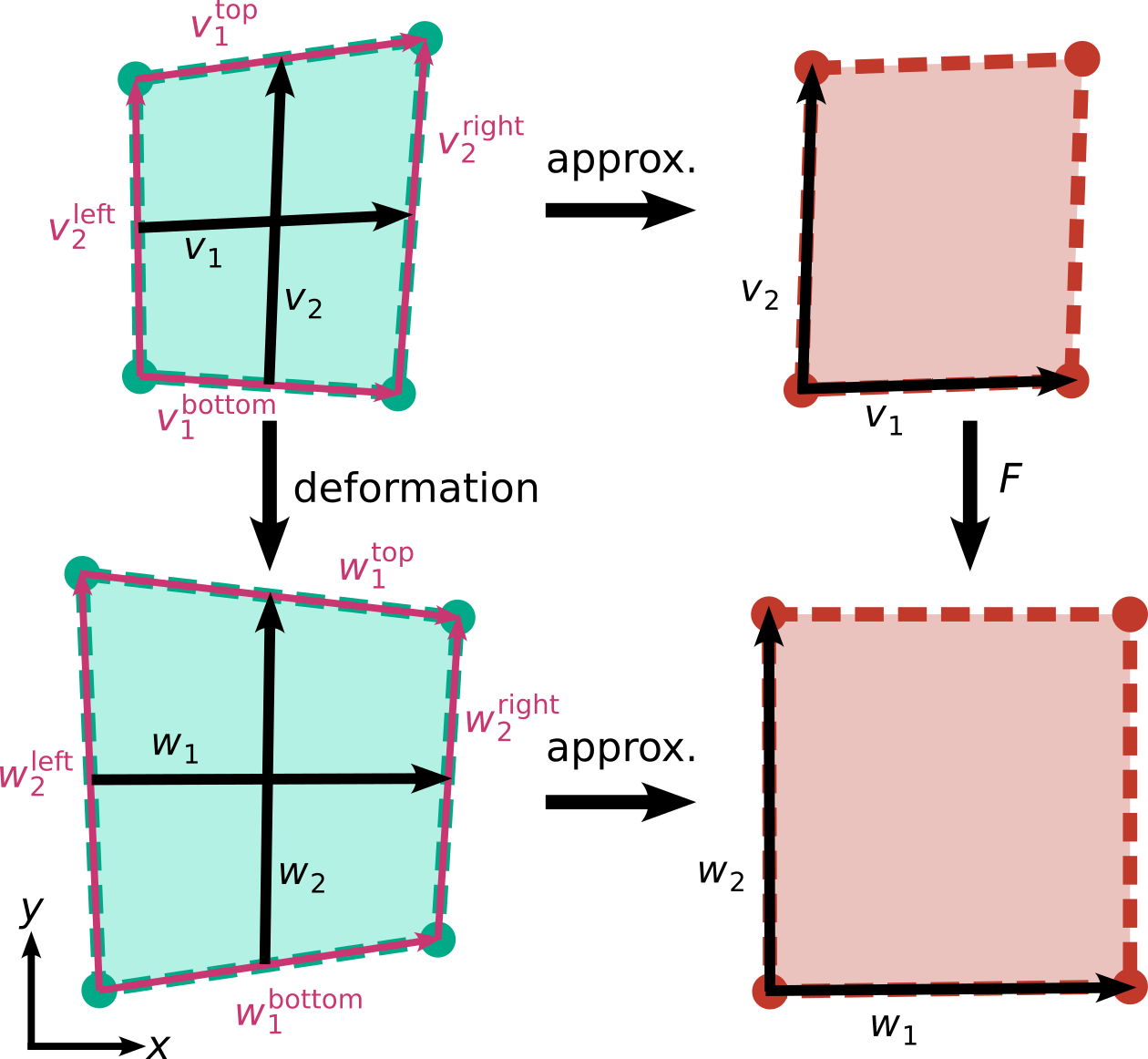}
    \caption{Schematic of the procedure used to calculate the lattice vectors in the reference $(\vec{v}_1,\,\vec{v}_2)$ and current $(\vec{w}_1,\, \vec{w}_2)$ configurations.}
    \label{fig:main_lattice_vecs}
\end{figure}

More specifically, for the $i$th quadrilateral defined by four tracked markers, we approximate it as a parallelogram by calculating the lattice vectors $\vec{v}_1^i$ and $\vec{v}_2^i$ in the undeformed configuration (first frame), and $\vec{w}_1^i$ and $\vec{w}_2^i$ in the deformed configuration (at a given frame). 
This approximation for a given quadrilateral is shown in Fig.~\ref{fig:main_lattice_vecs}.
By approximating it as a parallelogram, we know we have

\begin{equation}
    \begin{pmatrix}
        \vec{w}_1^i & \vec{w}_2^i
    \end{pmatrix}
    =
    \mathbf{F}^i
    \begin{pmatrix}
        \vec{v}_1^i & \vec{v}_2^i
    \end{pmatrix},
\end{equation}

where $\mathbf{F}^i$ is the deformation gradient for the $i$th quadrilateral, given by

\begin{equation} 
    \mathbf{F}^i = 
    \begin{pmatrix} 
    \diffp{{x^i}}{{X}} & \diffp{{x^i}}{{Y}}\\[1.5ex]
    \diffp{{y^i}}{{X}} & \diffp{{y^i}}{{Y}} 
    \end{pmatrix} = 
    \begin{pmatrix} 
    1 + \diffp{{u_x^i}}{{X}} & \diffp{{u_x^i}}{{Y}}\\[1.5ex]
    \diffp{{u_y^i}}{{X}} & 1 + \diffp{{u_y^i}}{{Y}} 
    \end{pmatrix}. 
\end{equation}

It follows that

\begin{equation}\label{eq:lattice_vec_eq}
   \begin{pmatrix} 
    1 + \diffp{{u_x^i}}{{X}} & \diffp{{u_x^i}}{{Y}}\\[1.5ex]
    \diffp{{u_y^i}}{{X}} & 1 + \diffp{{u_y^i}}{{Y}} 
    \end{pmatrix}
    =
    \begin{pmatrix}
        \vec{w}_1^i & \vec{w}_2^i
    \end{pmatrix}
    \begin{pmatrix}
        \vec{v}_1^i & \vec{v}_2^i
    \end{pmatrix}^{-1},
\end{equation}
from which, we compute the local normal strain components $\epsilon_{xx}^i = \diffp{{u_x^i}}{{X}}$ and $\epsilon_{yy}^i = \diffp{{u_y^i}}{{Y}}$ for each quadrilateral at each frame.

We also use our model to simulate the mechanical response of the knitted structures under tensile loading. To this end, we perform a virtual tensile test on the relaxed configuration by fixing the bottom boundary and imposing a prescribed displacement on the nodes along the top boundary.  Local strains are then computed using the same procedure employed in the experiments, enabling a direct comparison between simulations and measurements.

\begin{figure*}
  \includegraphics{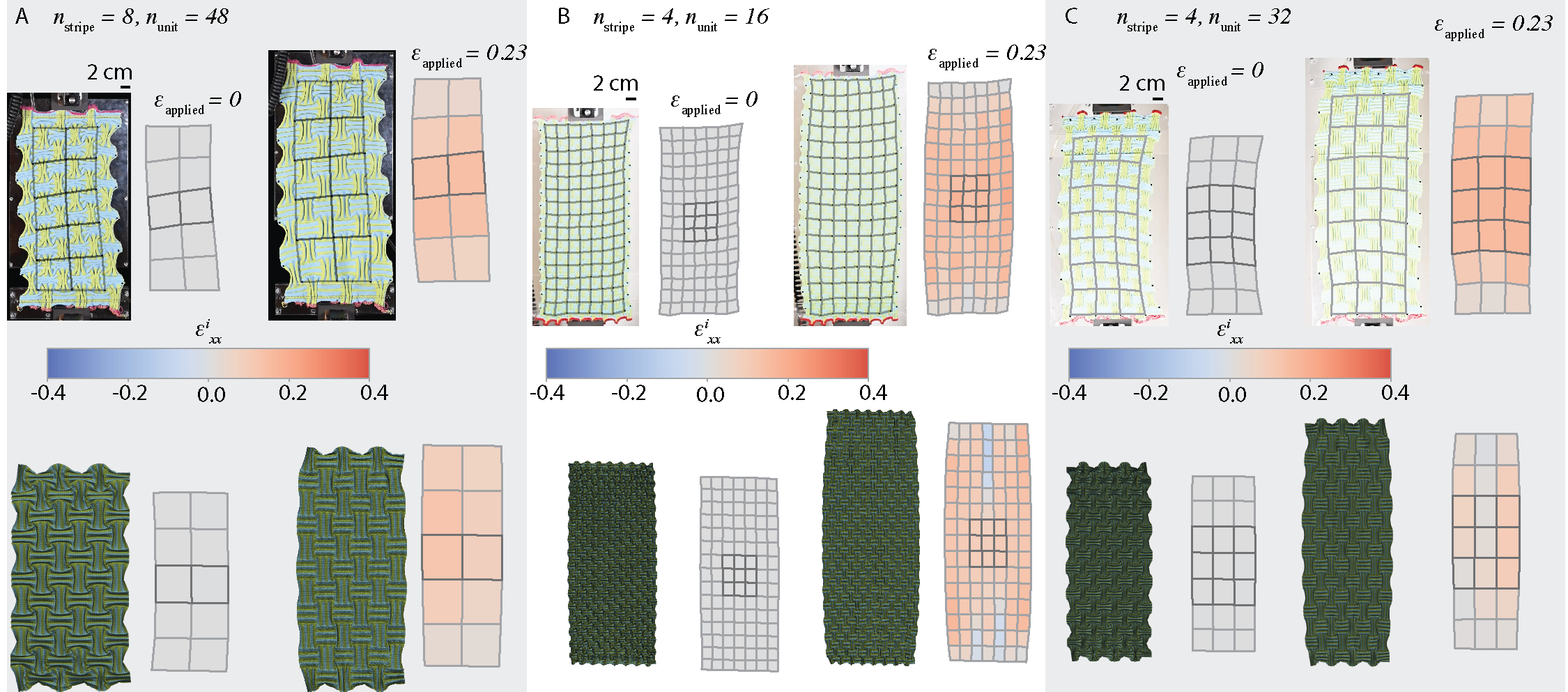}
  \caption{A--C) Relaxed ($\varepsilon_{\mathrm{applied}}=0$) and stretched ($\varepsilon_{\mathrm{applied}}=0.23$) configurations of fabrics composed of alternating rib and garter patches with $(n_{\text{stripe}}, n_{\text{subunit}})=$ A) $(8,48)$, B) $(4,16)$, and C) $(4,32)$. For each fabric, experimental results are shown in the top row and the corresponding numerical predictions in the bottom row. For both the experiments and simulations, we show a snapshot of the fabric (left) together with the corresponding map of the local transverse strain, $\varepsilon_{xx}^i$ (right).}
  \label{fig:tension_images}
\end{figure*}

In Fig.~\ref{fig:tension_images}, we plot the local strain, $\varepsilon_{xx}^i$, for samples with $(n_{\text{stripe}}, n_{\text{subunit}})=(8,48)$, $(4,16)$, and $(4,32)$, respectively, as measured experimentally (top row) and predicted numerically (bottom row) at different levels of applied global strain, $\varepsilon_{\text{applied}}$. Here, $\varepsilon_{\text{applied}}$ is defined as the displacement imposed at the sample boundaries divided by the initial sample length from clamp to clamp.
The local strain fields reveal the distribution of deformation within the sample and enable a direct comparison between the experimentally observed and numerically predicted kinematic responses.
Consistent across all samples, we find that the local transverse strain is positive and largest near the center of the textile, where the influence of the boundary conditions is weakest. Importantly, positive values of $\varepsilon_{xx}^i$ correspond to a negative Poisson's ratio, since the knits are subjected to tensile strain in the $y$-direction and expand, rather than contract, in the transverse $x$-direction.

\begin{figure*}
  \includegraphics{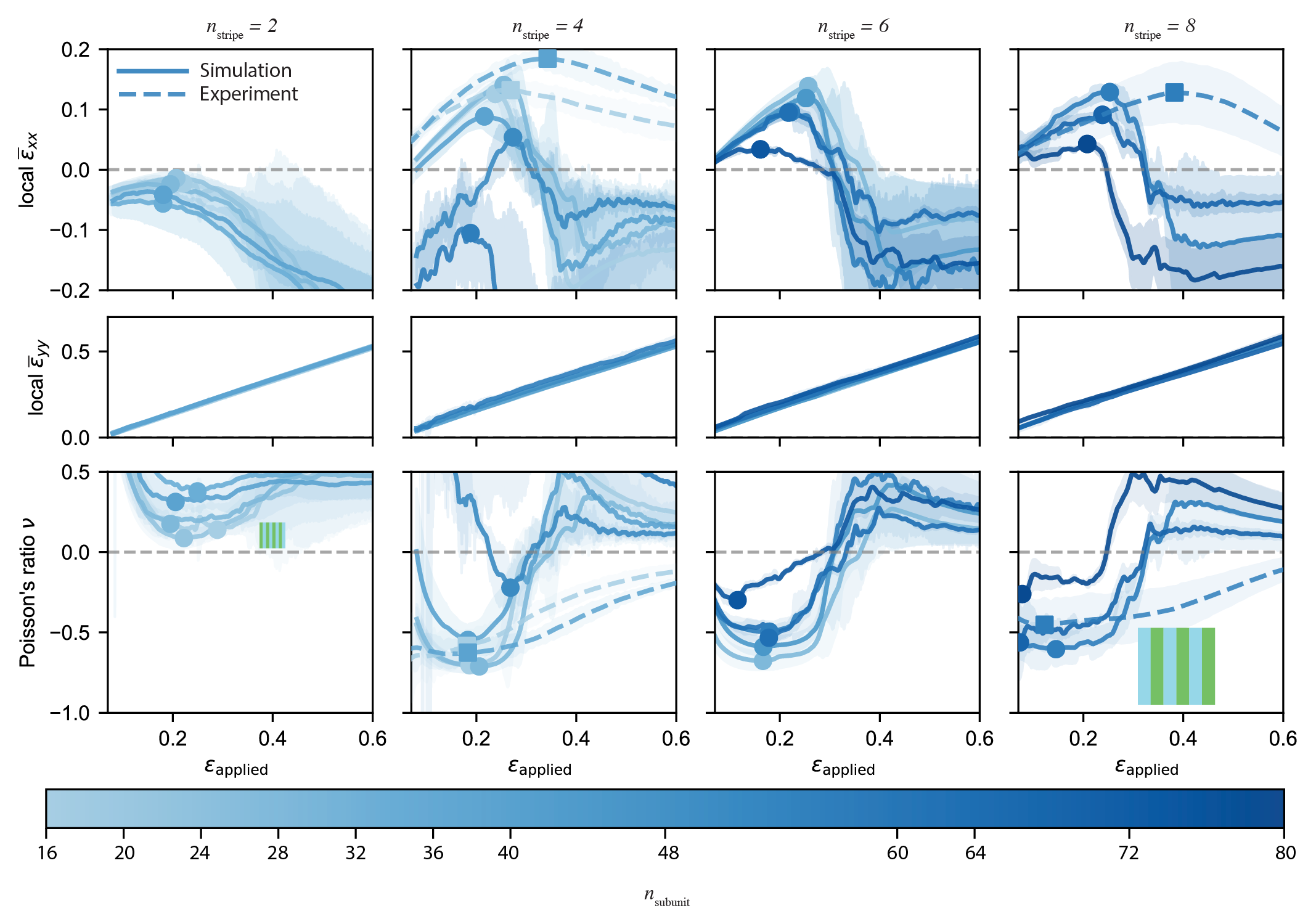}
  \caption{
 Evolution of the average local transverse strain, $\overline{\varepsilon}{xx}$ (top), the average local axial strain, $\overline{\varepsilon}{yy}$ (center), and the effective Poisson's ratio, $\nu$ (bottom), as functions of the applied strain, $\varepsilon_{\mathrm{applied}}$. Results are shown for fabrics with $n_{\text{stripe}}=2$, 4, 6, and 8. For each value of $n_{\text{stripe}}$, multiple values of $n_{\text{subunit}}$ are considered. Solid lines denote numerical predictions, while dashed lines represent experimental measurements.
  }  
  \label{fig:tension_plots}
\end{figure*}

Next, to quantitatively compare experiments and simulations, we calculate the average local strains, $\overline{\varepsilon}_{xx} = \langle \varepsilon_{xx}^i\rangle$ and $\overline{\varepsilon}_{yy} = \langle \varepsilon_{yy}^i\rangle$, over the central unit cells highlighted in dark gray in Fig.~\ref{fig:tension_images}, as functions of the applied strain. From these average strains, we compute the effective Poisson's ratio as
\begin{equation}
\nu=-\frac{\overline{\varepsilon}_{xx}}{\overline{\varepsilon}_{yy}}.
\end{equation}
The results reported in Fig.~\ref{fig:tension_plots} indicate that, for all three samples, $\overline{\varepsilon}_{xx}$ initially increases with the applied strain, reaches a maximum, and subsequently decreases. This behavior corresponds to the progressive unfolding of the sinusoidal boundaries and ridges within the knitted patches. The maximum transverse strain is attained once this unfolding process is complete. While both experiments and simulations exhibit the same qualitative trend, the decrease in $\overline{\varepsilon}_{xx}$ is more abrupt in the simulations. This discrepancy likely arises because the approximation of the rib and garter corrugations as a sequence of sharp folds becomes invalid once the textile is fully flattened under tension. At this stage, the simulated response is governed entirely by the linear spring network, whereas the deformation of the physical textile remains influenced by yarn mechanics and yarn--yarn interactions. As for $\overline{\varepsilon}_{yy}$, as expected, we find that it closely follows $\varepsilon_{\mathrm{applied}}$ and increases monotonically for all three samples. As a result, all three textiles exhibit an auxetic response, with the Poisson's ratio initially decreasing, reaching a minimum between $\varepsilon_\text{applied} = 0.1$ and $0.2$, and then progressively increasing as the applied strain is further increased. This non-monotonic evolution of $\nu$ directly reflects the unfolding-induced increase and subsequent decrease in $\overline{\varepsilon}_{xx}$.


We use the model to systematically investigate how $\nstripe$ and $n_\text{subunit}$ influence the evolution of the local strain and Poisson's ratio under uniaxial tension. In all cases, the relaxed configurations of the considered knits exhibit a combination of sinusoidal curvature and localized ridges. However, both the wavelength and amplitude of these corrugations depend on $\nstripe$ and $n_\text{subunit}$ (Fig.~\ref{fig:wavelength_analysis} for dependency of wavelength on $\nstripe$ and $n_\text{subunit}$). 
Upon stretching, the corrugations progressively unfold, causing the knitted structures to recover a flatter configuration.
The extent of this unfolding is primarily controlled by the stripe width. For $\nstripe \geq 4$, unfolding induces a lateral expansion in the $x$ direction, giving rise to auxetic behavior with minimum Poisson's ratios ranging from approximately $-0.2$ to $-0.7$. 
In contrast, for $\nstripe = 2$, the ridges and sinusoidal boundaries are much less pronounced, resulting in a substantially smaller unfolding regime (Fig.~\ref{fig: rib x garter}D--E). 
Consequently, stretching produces little transverse expansion, and none of these unit cells exhibit auxetic behavior; instead, their Poisson's ratios remain slightly positive. For a fixed stripe width, $n_\text{subunit}$ further modulates the mechanical response. Increasing $n_\text{subunit}$ systematically reduces the average transverse strain, $\overline{\varepsilon}_{xx}$, thereby increasing the minimum Poisson's ratio (i.e., making the response less auxetic). 
For example, when $\nstripe = 4$, unit cells with $\nunit \leq 40$ exhibit auxetic behavior, while the unit cell with $\nunit = 48$ does not.
Overall, wider stripes and smaller unit cells both promote stronger auxeticity. 
Among the geometries considered, the unit cell with $n_\text{stripe}=4$ and $n_\text{subunit}=16$ exhibits the most negative Poisson's ratio. 
Importantly, varying the unit-cell geometry affects not only the magnitude of the minimum Poisson's ratio but also the applied strain at which it is attained. 
This tunability enables the design of knitted metamaterials that exhibit a prescribed auxetic response at a desired deformation level, providing an additional degree of freedom for programming shape change.

\begin{figure}
  \includegraphics{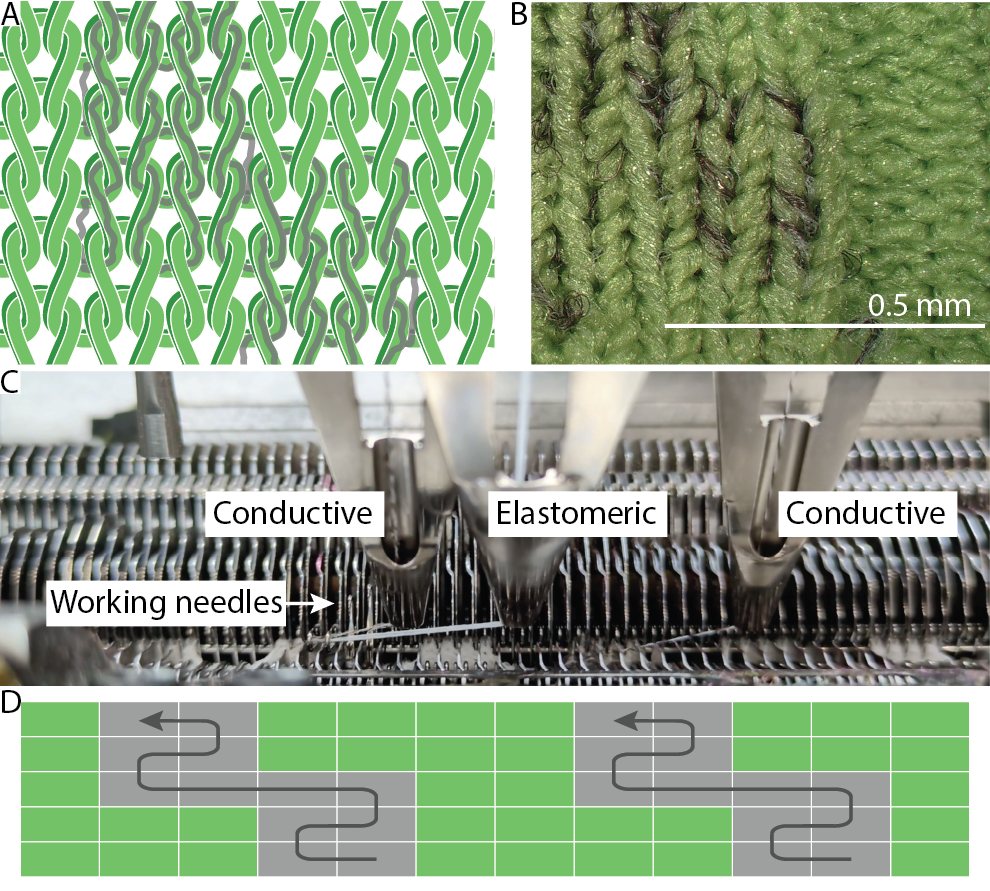}
  \caption{A) Schematic illustrating the \textit{partial plating} technique, in which a conductive yarn is selectively plated with inert yarns, enabling arbitrary conductive pathways to be embedded within the textile. B) Photograph of a sample fabricated using the \textit{partial plating} technique. C) Photograph of the knitting machine. Active needles knit both the conductive and elastomeric yarns to create the patterned conductive pathways that form the capacitors. D) Examples of   arbitrary conductive pathways possible using the \textit{partial plating} technique.}
  \label{fig: capacitor fabrication}
\end{figure}
\begin{figure}
  \includegraphics{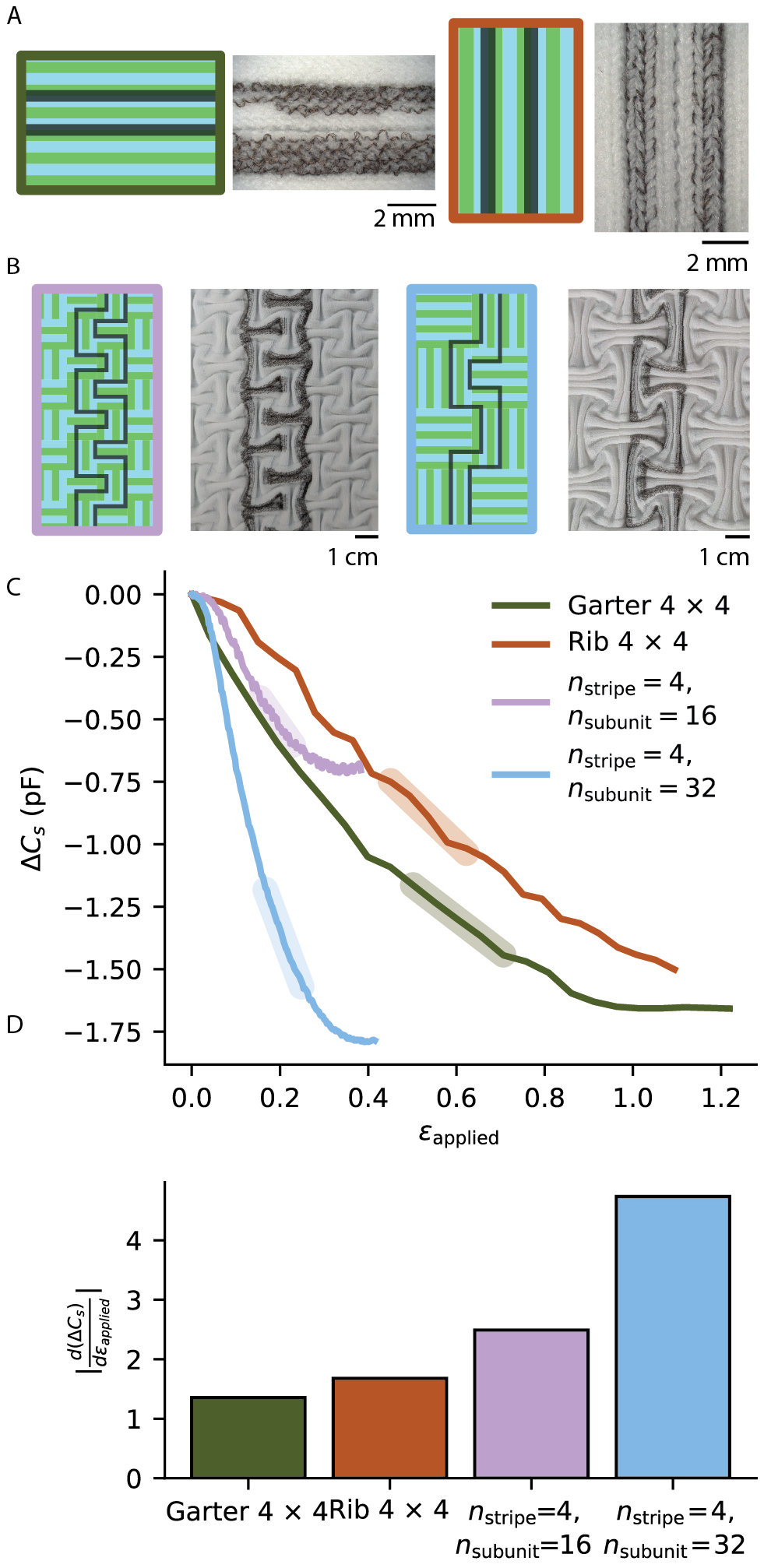}
  \caption{A--B) Using the partial plating technique, we knit conductive pathways into A) $4 \times 4$ garter and rib fabrics and B) auxetic fabrics with $(n_{\text{stripe}}, n_{\text{subunit}})=(4,16)$ and $(4,32)$. C) Measured capacitance as a function of the applied strain for the four fabrics. D) Average rate of capacitance change computed over the central 20\% of the measured strain range (highlighted by the shaded regions in panel C) for each of the four fabrics. }
  \label{fig: capacitor testing}
\end{figure}

\section{Knit capacitors}
Previous work has demonstrated textile sensors in the form of resistors and capacitors as strain gauges and pressure sensors \cite{atalay2017batch,sanchez2023knitting, abbas_Tuning_2026}. Here we demonstrate a novel method of creating textile capacitors in a single manufacturing step. 
When we knit a stitch using multiple yarn carriers, one yarn lies on the face of the textile while one lies on the reverse. This technique is referred to as \textit{plated knitting}. One use of plated knitting is to highlight the knitting pattern by using two different colors, as stitches made on the front bed will be a different color than stitches made on the back bed (Fig. \ref{fig: rib x garter}A) \cite{mahadevan2024knitting}.

To knit pre-programmed conductive patches, we plate a fine conductive yarn alongside the elastomeric yarns for a few select stitches in each row (Fig. \ref{fig: capacitor fabrication}A-B); we refer to this technique as \textit{partial plating}. 
Each group of stitches (conductive or not), must be knit in its own pass \cite{sanchez2023knitting}; this splits each row into multiple knitting operations (ref SI video).
By incorporating multiple conductive yarns in a single row of knitting, we can create arbitrary conductive pathways and patches (Fig. \ref{fig: capacitor fabrication}C-D).
In Figs.~\ref{fig: capacitor testing}A--B, we present four samples with embedded conductive pathways. 
First, we consider garter and rib knits and knit two parallel conductive pathways along their respective ridges (Fig.~\ref{fig: capacitor testing}A), thereby creating parallel-plate capacitors. Next, we fabricate curved, non-parallel-plate capacitors by plating conductive yarns onto the ridges of fabrics composed of alternating rib and garter patches. This is shown in  Fig.~\ref{fig: capacitor testing}B for knits with $(n_\text{stripe}, n_\text{subunit})=(4,16)$ and $(4,32)$.

For each sample, we apply a prescribed displacement using an Instron tensile tester while simultaneously measuring the capacitance with a benchtop LCR meter (Keysight E4980AL). The mechanical and electrical measurements are synchronized using a voltage pulse sent to the Instron, and the capacitance is continuously recorded at an interrogation frequency of 1000 Hz. To focus on the unfolding regime, each sample is stretched only until its ridges are fully unfolded. Consequently, the maximum applied strain differs between samples.
Fig.~\ref{fig: capacitor testing}C reports the measured capacitance as a function of applied strain for rib, garter, and two knitted fabrics composed of alternating rib and garter patches. The rib knit exhibits the smallest sensitivity, with its capacitance decreasing slowly as the applied strain increases. In contrast, the knit with $(n_\text{stripe},n_\text{subunit})=(4,32)$ exhibits the steepest decrease in capacitance, corresponding to the highest strain sensitivity. However, this enhanced sensitivity is confined to the strain range of approximately $0.1\leq\varepsilon\leq0.3$, after which the unfolding process is complete and the capacitance changes much more slowly. As a result, this design is well suited for sensing relatively small strains but offers a limited operating range.

To quantitatively compare the different designs, we compute the average rate of capacitance change over the central 20\% of the measured strain range (highlighted by the shaded areas in Fig.~\ref{fig: capacitor testing}C). The magnitude of this average rate of capacitance change  is reported in Fig.~\ref{fig: capacitor testing}D. Together, the results of Figs.~\ref{fig: capacitor testing}C-D demonstrate that the knitted capacitor architecture can be tuned to realize either highly sensitive strain sensors operating over a narrow strain window or less sensitive sensors capable of measuring substantially larger strains. In Fig.~\ref{fig:tension_plots}, we have laid out an entire family of potential sensors; we predict that by varying $n_\text{stripe}$ and $n_\text{subunit}$, we will be able to tune both the slope of the strain-capacitance relationship and the strain range over which this slope is constant and therefore design a sensing response tuned to a particular structure. 

\section{Conclusion}

In summary, we investigate textiles composed of checkerboard arrangements of rib and garter patches that give rise to complex three-dimensional corrugated geometries. To capture their nonlinear mechanical behavior, we develop a reduced-order modeling framework based on a network of linear torsional and extensional springs. We demonstrated that the model accurately predicts both the geometry and kinematic response of these textiles. Combining simulations and experiments, we systematically characterized how the geometric parameters of the checkerboard pattern influence the mechanical response. We found that tensile loading induces pronounced internal rotations that generate auxetic behavior, with the magnitude and onset of the negative Poisson's ratio governed by the textile architecture. Finally, we integrated capacitive sensing functionality into the fabrics and showed how their initial corrugated configurations and deformation kinematics influence sensing performance. Motivated by prior work on auxetic capacitive sensors \cite{dorsey2022origami}, we anticipate that the rotations inherent to auxetic designs can enhance strain-sensing performance. Consistent with this expectation, we observe a steeper capacitance--strain response for the auxetic unit cell, although over a narrower strain range.

Several extensions could  improve the accuracy and applicability of the spring-network model. For example, the spring stiffnesses could be fitted directly to the kinematic or mechanical response of the full auxetic textile, rather than to the isolated rib and garter building blocks. Incorporating nonlinear springs could further improve agreement with experiments over larger strain ranges. Finally, using a denser mesh to better capture the cross-sectional curvature of rib and garter stitches could increase model fidelity, albeit at the cost of additional fitting parameters and computational expense. Looking forward, this spring-network framework could also be used as a design tool to target prescribed shapes, tailored auxetic responses, or optimized sensing and mechanical properties.

While we have focused here on textiles composed of checkerboard arrangements of rib and garter patches, the combined experimental and numerical framework developed in this work could be extended to a much wider class of knitted architectures. In particular, it could be applied to knit structures that more closely resemble traditional origami, with well-defined flat faces and localized folds, as recently demonstrated in textile systems \cite{salmon2025structural}. Although such structures may be less conventional as wearable textiles, they could provide more predictable and programmable mechanical responses. The ability to capture geometry, kinematics, and mechanical response within a unified reduced-order framework offers a promising route toward the rational design of knitted materials with tailored shape-morphing, mechanical, and sensing functionalities.
\section*{Methods}
All knit samples in this work were produced on a Shima Seiki SWG N2 061 machine. All samples were knit at a stitch size of `34' in machine units; the conductive patches were knit at a stitch size of `38.' 
All rib samples tested were 196 stitches wide by 192 stitches long, and all garter samples were 92 stitches wide by 384 stitches long. Rib and garter samples were tested on hook grips \cite{sanchez2023knitting,mahadevan2024knitting} and loaded at 40 mm/min. Combined rib and garter samples were loaded on a custom rig that combined the hook grips with acrylic plates to prevent out-of-plane buckling. 
All samples were knit with two ends of Elastomeric Nylon Lycra (Yeoman Yarns, United Kingdom), made from 81\% nylon and 19\% Lycra. The conductive yarn used is `20 dtex Lycra + 44Ag + 44 roh Twisted Yarn', purchased from V Technical Textiles, Palmyra, NY. 

\section*{Data Availability Statement}
The data that support the findings of this study are openly available on \href{https://github.com/bertoldi-collab/textile_folding_simulations}{github}.

\section*{Acknowledgments}
Research was supported by the Simons Collaboration on Extreme Wave Phenomena Based on Symmetries. Equipment was supported by ONR DURIP Award N00014-19-1-2220.  The authors would like to thank Dr. Giovanni Bordiga and Dr. Anne Meeussen for insightful conversations and feedback.
\printbibliography[title={References}]

@misc{bordiga2023humble,
    title        = {A humble image tracking code},
    author       = {Bordiga, Giovanni},
    year         = {2023},
    howpublished = {\url{https://github.com/bertoldi-collab/tracking-markers}},
    note         = {Bertoldi Group}
}

@article{shan2015design,
  title={Design of planar isotropic negative Poisson’s ratio structures},
  author={Shan, Sicong and Kang, Sung H and Zhao, Zhenhao and Fang, Lichen and Bertoldi, Katia},
  journal={Extreme Mechanics Letters},
  volume={4},
  pages={96--102},
  year={2015},
  publisher={Elsevier}
}

@article{abbas_Tuning_2026,
	title = {Tuning the {Electromechanical} {Performance} of {Knitted} {Strain} {Sensors} through {Stitch} {Variation}},
	volume = {4},
	url = {https://doi.org/10.1021/acsaenm.5c00925},
	doi = {10.1021/acsaenm.5c00925},
	number = {1},
	urldate = {2026-03-22},
	journal = {ACS Applied Engineering Materials},
	publisher = {American Chemical Society},
	author = {Abbas, Adeel and Choudhry, Nauman Ali and Underwood, Jenny and Houshyar, Shadi and Wang, Xin},
	month = jan,
	year = {2026},
	pages = {243--253},
}

@article{dorsey2022origami,
  title={Origami-patterned capacitor with programmed strain sensitivity},
  author={Dorsey, Kristen L and Huang, HuiYing and Wen, Yuhan},
  journal={Multifunctional Materials},
  volume={5},
  number={2},
  pages={025001},
  year={2022},
  publisher={IOP Publishing}
}

@book{salmon2025structural,
  title={Structural Stitches: A Machine Knitter's Guide to Creating Form and Structure},
  author={Victoria Salmon},
  year={2025},
  publisher={The Crowood Press Ltd}
}

@inproceedings{kilic2021omnifiber,
  title={OmniFiber: Integrated fluidic fiber actuators for weaving movement based interactions into the ‘fabric of everyday life’},
  author={Kilic Afsar, Ozgun and Shtarbanov, Ali and Mor, Hila and Nakagaki, Ken and Forman, Jack and Modrei, Karen and Jeong, Seung Hee and Hjort, Klas and H{\"o}{\"o}k, Kristina and Ishii, Hiroshi},
  booktitle={The 34th Annual ACM Symposium on User Interface Software and Technology},
  pages={1010--1026},
  year={2021}
}

@article{ma2017review,
  title={Review on the knitted structures with auxetic effect},
  author={Ma, Pibo and Chang, Yuping and Boakye, Andrews and Jiang, Gaoming},
  journal={The Journal of The Textile Institute},
  volume={108},
  number={6},
  pages={947--961},
  year={2017},
  publisher={Taylor \& Francis}
}

@article{o2022unfolding,
  title={Unfolding textile-based pneumatic actuators for wearable applications},
  author={O'Neill, Ciar{\'a}n T and McCann, Connor M and Hohimer, Cameron J and Bertoldi, Katia and Walsh, Conor J},
  journal={Soft Robotics},
  volume={9},
  number={1},
  pages={163--172},
  year={2022},
  publisher={SAGE Publications Sage CA: Los Angeles, CA}
}

@article{nguyen2020design,
  title={Design and computational modeling of fabric soft pneumatic actuators for wearable assistive devices},
  author={Nguyen, Pham Huy and Zhang, Wenlong},
  journal={Scientific reports},
  volume={10},
  number={1},
  pages={9638},
  year={2020},
  publisher={Nature Publishing Group UK London}
}

@article{du2024haptiknit,
  title={Haptiknit: Distributed stiffness knitting for wearable haptics},
  author={du Pasquier, Cosima and Tessmer, Lavender and Scholl, Ian and Tilton, Liana and Chen, Tian and Tibbits, Skylar and Okamura, Allison},
  journal={Science Robotics},
  volume={9},
  number={97},
  pages={eado3887},
  year={2024},
  publisher={American Association for the Advancement of Science}
}

@article{eschen2019performance,
  title={Performance and prediction of large deformation contractile shape memory alloy knitted actuators},
  author={Eschen, Kevin and Abel, Julianna},
  journal={Smart Materials and Structures},
  volume={28},
  number={2},
  pages={025014},
  year={2019},
  publisher={IOP Publishing}
}

@article{mahadevan2024knitting,
  title={Knitting multistability},
  author={Mahadevan, Kausalya and Yuen, Michelle C and Farrell, David T and Walsh, Conor J and Sanchez, Vanessa and Wood, Robert J and Bertoldi, Katia},
  journal={Advanced Functional Materials},
  pages={e76385},
  year={2026},
  publisher={Wiley Online Library}
}

@article{tajiri2025curling,
  title={Curling morphology of knitted fabrics: Structure and Mechanics},
  author={Tajiri, Kotone and Murakami, Riki and Kobayashi, Shunsuke and Tarumi, Ryuichi and Sano, Tomohiko G},
  journal={Extreme Mechanics Letters},
  volume={76},
  pages={102300},
  year={2025},
  publisher={Elsevier}
}

@article{niu2025geometric,
  title={Geometric modeling of knitted fabrics},
  author={Niu, Lauren and Dion, Genevi{\`e}ve and Kamien, Randall D},
  journal={Proceedings of the National Academy of Sciences},
  volume={122},
  number={7},
  pages={e2416536122},
  year={2025},
  publisher={National Academy of Sciences}
}

@article{ding2024unravelling,
  title={Unravelling the mechanics of knitted fabrics through hierarchical geometric representation},
  author={Ding, Xiaoxiao and Sanchez, Vanessa and Bertoldi, Katia and Rycroft, Chris H},
  journal={Proceedings of the Royal Society A},
  volume={480},
  number={2295},
  pages={20230753},
  year={2024},
  publisher={The Royal Society}
}

@article{Atalay2018ALayer,
    title = {{A Highly Sensitive Capacitive-Based Soft Pressure Sensor Based on a Conductive Fabric and a Microporous Dielectric Layer}},
    year = {2018},
    journal = {Advanced Materials Technologies},
    author = {Atalay, Ozgur and Atalay, Asli and Gafford, Joshua and Walsh, Conor},
    number = {1},
    pages = {1--8},
    volume = {3},
    doi = {10.1002/admt.201700237},
    issn = {2365709X}
}

@article{Narayanan2018AutomaticMeshes,
    title = {{Automatic machine knitting of 3d meshes}},
    year = {2018},
    journal = {ACM Transactions on Graphics},
    author = {Narayanan, Vidya and Albaugh, Lea and Hodgins, Jessica and Coros, Stelian and McCann, James},
    number = {3},
    volume = {37},
    doi = {10.1145/3186265},
    issn = {15577368}
}

@article{Hu2019AuxeticTextiles,
    title = {{Auxetic textiles}},
    year = {2019},
    journal = {Auxetic Textiles},
    author = {Hu, Hong and Zhang, Minglonghai and Liu, Yanping},
    pages = {1--358},
    isbn = {9780081022115},
    doi = {10.1016/C2016-0-04399-1},
    issn = {1318-0207},
    pmid = {24362973}
}

@article{Alderson2012AuxeticStructures,
    title = {{Auxetic warp knit textile structures}},
    year = {2012},
    journal = {Physica Status Solidi (B) Basic Research},
    author = {Alderson, Kim and Alderson, Andrew and Anand, Subhash and Simkins, Virginia and Nazare, Shonali and Ravirala, Naveen},
    number = {7},
    pages = {1322--1329},
    volume = {249},
    doi = {10.1002/pssb.201084216},
    issn = {03701972}
}

@article{Hu2011DevelopmentTechnology,
    title = {{Development of auxetic fabrics using flat knitting technology}},
    year = {2011},
    journal = {Textile Research Journal},
    author = {Hu, Hong and Wang, Zhengyue and Liu, Su},
    number = {14},
    pages = {1493--1502},
    volume = {81},
    isbn = {0040-5175},
    doi = {10.1177/0040517511404594},
    issn = {00405175}
}

@inproceedings{Kaldor2008SimulatingLevel,
    title = {{Simulating cloth at the yarn level}},
    year = {2008},
    author = {Kaldor, Jonathan},
    pages = {100--100},
    doi = {10.1145/1400468.1400541}
}

@article{granberry2019functionally,
  title={Functionally graded knitted actuators with NiTi-based Shape Memory Alloys for topographically self-fitting wearables},
  author={Granberry, Rachael and Eschen, Kevin and Holschuh, Brad and Abel, Julianna},
  journal={Advanced materials technologies},
  volume={4},
  number={11},
  pages={1900548},
  year={2019},
  publisher={Wiley Online Library}
}

@article{singal2024programming,
  title={Programming mechanics in knitted materials, stitch by stitch},
  author={Singal, Krishma and Dimitriyev, Michael S and Gonzalez, Sarah E and Cachine, A Patrick and Quinn, Sam and Matsumoto, Elisabetta A},
  journal={Nature Communications},
  volume={15},
  number={1},
  pages={2622},
  year={2024},
  publisher={Nature Publishing Group UK London}
}

@article{sanchez2023knitting,
  title={3D Knitting for Pneumatic Soft Robotics},
  author={Sanchez, Vanessa and Mahadevan, Kausalya and Ohlson, Gabrielle and Graule, Moritz A and Yuen, Michelle C and Teeple, Clark B and Weaver, James C and McCann, James and Bertoldi, Katia and Wood, Robert J},
  journal={Advanced Functional Materials},
  pages={2212541},
  year={2023},
  publisher={Wiley Online Library}
}

@inproceedings{elmoughni2021machine,
  title={Machine-knitted seamless pneumatic actuators for soft robotics: design, fabrication, and characterization},
  author={Elmoughni, Hend M and Yilmaz, Ayse Feyza and Ozlem, Kadir and Khalilbayli, Fidan and Cappello, Leonardo and Tuncay Atalay, Asli and Ince, G{\"o}khan and Atalay, Ozgur},
  booktitle={Actuators},
  volume={10},
  number={5},
  pages={94},
  year={2021},
  organization={MDPI}
}

@inproceedings{luo2022digital,
  title={Digital fabrication of pneumatic actuators with integrated sensing by machine knitting},
  author={Luo, Yiyue and Wu, Kui and Spielberg, Andrew and Foshey, Michael and Rus, Daniela and Palacios, Tom{\'a}s and Matusik, Wojciech},
  booktitle={Proceedings of the 2022 CHI Conference on Human Factors in Computing Systems},
  pages={1--13},
  year={2022}
}

@article{atalay2017batch,
  title={Batch fabrication of customizable silicone-textile composite capacitive strain sensors for human motion tracking},
  author={Atalay, Asli and Sanchez, Vanessa and Atalay, Ozgur and Vogt, Daniel M and Haufe, Florian and Wood, Robert J and Walsh, Conor J},
  journal={Advanced Materials Technologies},
  volume={2},
  number={9},
  pages={1700136},
  year={2017},
  publisher={Wiley Online Library}
}

@article{sanchez2021textile,
  title={Textile technology for soft robotic and autonomous garments},
  author={Sanchez, Vanessa and Walsh, Conor J and Wood, Robert J},
  journal={Advanced functional materials},
  volume={31},
  number={6},
  pages={2008278},
  year={2021},
  publisher={Wiley Online Library}
}

@article{granberry2021kinetically,
  title={Kinetically tunable, active auxetic, and variable recruitment active textiles from hierarchical assemblies},
  author={Granberry, Rachael and Barry, Justin and Holschuh, Brad and Abel, Julianna},
  journal={Advanced Materials Technologies},
  volume={6},
  number={3},
  pages={2000825},
  year={2021},
  publisher={Wiley Online Library}
}

@phdthesis{Underwood2009Fix,
author = {Underwood, Jenny},
school = {RMIT University},
title = {{The Design of 3D Shape Knitted Preforms}},
type = {Doctoral Thesis},
year = {2009}
}

@inproceedings{Granberry2017activeorthostaticfix,
author = {Granberry, Rachael and Abel, Julianna and Holschuh, Brad},
title = {Active Knit Compression Stockings for the Treatment of Orthostatic Hypotension},
year = {2017},
isbn = {9781450351881},
publisher = {Association for Computing Machinery},
address = {New York, NY, USA},
url = {https://doi.org/10.1145/3123021.3123065},
doi = {10.1145/3123021.3123065},
booktitle = {Proceedings of the 2017 ACM International Symposium on Wearable Computers},
pages = {186–191},
numpages = {6},
location = {Maui, Hawaii},
series = {ISWC '17}
}

@article{Eschen2020materialia,
author = {Eschen, Kevin and Garcia-Barriocanal, Javier and Abel, Julianna},
doi = {10.1016/j.mtla.2020.100684},
issn = {25891529},
journal = {Materialia},
pages = {100684},
publisher = {Elsevier B.V.},
title = {{In-situ strain- and temperature-control X-ray micro-diffraction analysis of nickel–titanium knitted architectures}},
url = {https://doi.org/10.1016/j.mtla.2020.100684},
volume = {11},
year = {2020}
}

\end{document}